# Discovery of two-dimensional anisotropic superconductivity at KTaO$_3$ (111) interfaces

**Authors:** Changjiang Liu[1*†], Xi Yan[1,2,5 †], Dafei Jin[3], Yang Ma[6], Haw-Wen Hsiao[4], Yulin Lin[3], Terence M. Bretz-Sullivan[1], Xianjing Zhou[3], John Pearson[1], Brandon Fisher[3], J. Samuel Jiang[1], Wei Han[6], Jian-Min Zuo[4], Jianguo Wen[3], Dillon D. Fong[1], Jirong Sun[5], Hua Zhou[2], Anand Bhattacharya[1*]

**Affiliations:**

[1]Materials Science Division, Argonne National Laboratory, Lemont, IL 60439, USA.

[2]Advanced Photon Source, Argonne National Laboratory, Lemont, IL 60439, USA.

[3]Nanoscale Science and Technology Division, Argonne National Laboratory, Lemont, IL 60439, USA.

[4]Department of Materials Science and Engineering, University of Illinois at Urbana-Champaign, Urbana, Illinois 61801, USA.

[5]Beijing National Laboratory for Condensed Matter & Institute of Physics, Chinese Academy of Sciences, Beijing 100190, People's Republic of China.

[6]International Centre for Quantum Materials, School of Physics, Peking University, Beijing 100871, People's Republic of China.

*Correspondence to anand@anl.gov or changjiang.liu@anl.gov

†These authors contributed equally.

**Abstract:** The unique electronic structure found at interfaces between materials can allow unconventional quantum states to emerge. Here we observe superconductivity in electron gases formed at interfaces between (111) oriented KTaO$_3$ and insulating overlayers of either EuO or LaAlO$_3$. The superconducting transition temperature, approaching 2.2 K, is about one order of magnitude higher than that of the LaAlO$_3$/SrTiO$_3$ system. Strikingly, similar electron gases at (001) KTaO$_3$ interfaces remain normal down to 25 mK. The critical field and current-voltage measurements indicate that the superconductivity is two dimensional. Higher mobility EuO/KTaO$_3$ (111) samples show a large in-plane anisotropy in transport properties at low temperatures prior to onset of superconductivity, suggesting the emergence of a 'stripe' like phase where the superconductivity is nearly homogeneous in one direction, but strongly modulated in the other.

**One Sentence Summary:** Electron gases formed at KTaO$_3$ (111) interfaces show 2D anisotropic superconductivity with $T_c$ much higher than that in LaAlO$_3$/SrTiO$_3$.

**Main Text:**

## Introduction

Superconductivity in two dimensions (2D) has been a central theme in condensed matter physics and material science (*1*). In 2D, the electron-electron and electron-lattice interactions that mediate pairing can also give rise to states that compete with superconductivity, and fluctuations can further suppress 2D superconductivity. Thus, while there are many examples of two-dimensional electron gases (2DEG) and ultrathin metallic films, only a small fraction of these are superconducting. Most of the foundational work in 2D superconductivity was carried out using amorphous thin films (*2*), and led to deep insights regarding the nature of classical and quantum phase transitions in the 2D limit. In more recent years, there have been striking developments in the realization of 2D superconductivity in *crystalline* materials (*1*), including layered transition metal chalcogenides (*3, 4*), bilayer (*5*) and trilayer (*6*) graphene, gated surfaces of single crystals (*7, 8*), atomically thin ordered metallic layers on semiconductor surfaces (*9*), and interfaces between crystalline materials (*10-12*). These crystalline 2D superconductors allow us to realize and break symmetries, and tailor electronic structure in ways that are not possible in amorphous and disordered thin films. For example, in a 2D superconductor with strong spin-orbit coupling and broken inversion symmetry, a Rashba interaction may lead to a superconducting order parameter with mixed *s*-wave and *p*-wave symmetry (*13*), a candidate platform for realizing Majorana modes (*14*). It is perhaps telling that three of the most prominent examples of 2D superconductivity at crystalline interfaces – $LaAlO_3/SrTiO_3$ (LAO/STO), $FeSe/SrTiO_3$ and $La_2CuO_4/LaSrCuO_4$ - involve transition metal oxides (*10-12*), where strong electron-electron and electron-lattice interactions likely play a role in mediating superconducting pairing. In some of these interfacial systems, the superconductivity is believed to be 'unconventional' – i.e. with a non *s*-wave order parameter, and a pairing mechanism that is not described by the Bardeen-Cooper-Schrieffer (BCS) theory.

$KTaO_3$ (KTO) is an insulator with a cubic perovskite structure, a bandgap of 3.6 eV, and a dielectric constant that grows to values > 4500 upon cooling to low temperatures (*15, 16*). It is believed that KTO is on the verge of a ferroelectric transition, thwarted by quantum fluctuations at low temperatures. It is thus called a 'quantum paraelectric', much like $SrTiO_3$ (STO). STO can be easily doped into a metallic state that becomes superconducting at $T < 0.5$ K (*17*). In contrast, few studies on doped KTO exist for bulk samples (*18, 19*), and none of them report any signatures of superconductivity. Nonetheless, it was reported that ionic liquid gating could be used to tune the KTO (001) surface into a weak superconducting state with $T_c \sim 47$ mK (*8*). The electronic states near the Fermi level in electron doped KTO are derived from Ta-5$d$ states (*20*) – in particular $d_{xy}$, $d_{zx}$ and $d_{yz}$ orbitals in the $t_{2g}$ manifold. The large spin-orbit coupling in these states lifts the degeneracy of the $t_{2g}$ manifold and splits the $J = 1/2$ states to an energy $\sim 0.4$ eV higher than the $J = 3/2$ states (compared to $\sim 0.025$ eV in STO). The two $J = 3/2$ states ($m_J = \pm 1/2$ and $m_J = \pm 3/2$) are degenerate near the Γ-point and disperse as 'light' and 'heavy' electron states at finite momenta. Unlike in STO, there are no symmetry lowering lattice distortions at low temperatures to lift this degeneracy at the Γ-point, though confinement can also lift degeneracies. 2DEGs can be realized at the KTO (001) and (111) interfaces via vacuum cleaving (*21, 22*) followed by exposure to UV or synchrotron radiation, and have been characterized using angle resolved photoemission spectroscopy (ARPES) (*21*). ARPES studies on KTO (111) surface found a distinct Fermi surface with the six-fold symmetry of the surface lattice (Fig. 1a,b), spin split-off bands, and evidence for band bending and confinement near the surface (*21, 23*). Owing to the large spin-orbit coupling, the interfacial electron gas at KTO (111) interfaces (*24, 25*) is also expected to host complex spin textures in momentum space. In the particular case of a bilayer of KTO along the [111] direction, which forms a puckered honeycomb lattice (Fig. 1b), calculations suggest that this system may support topological states (*26, 27*).

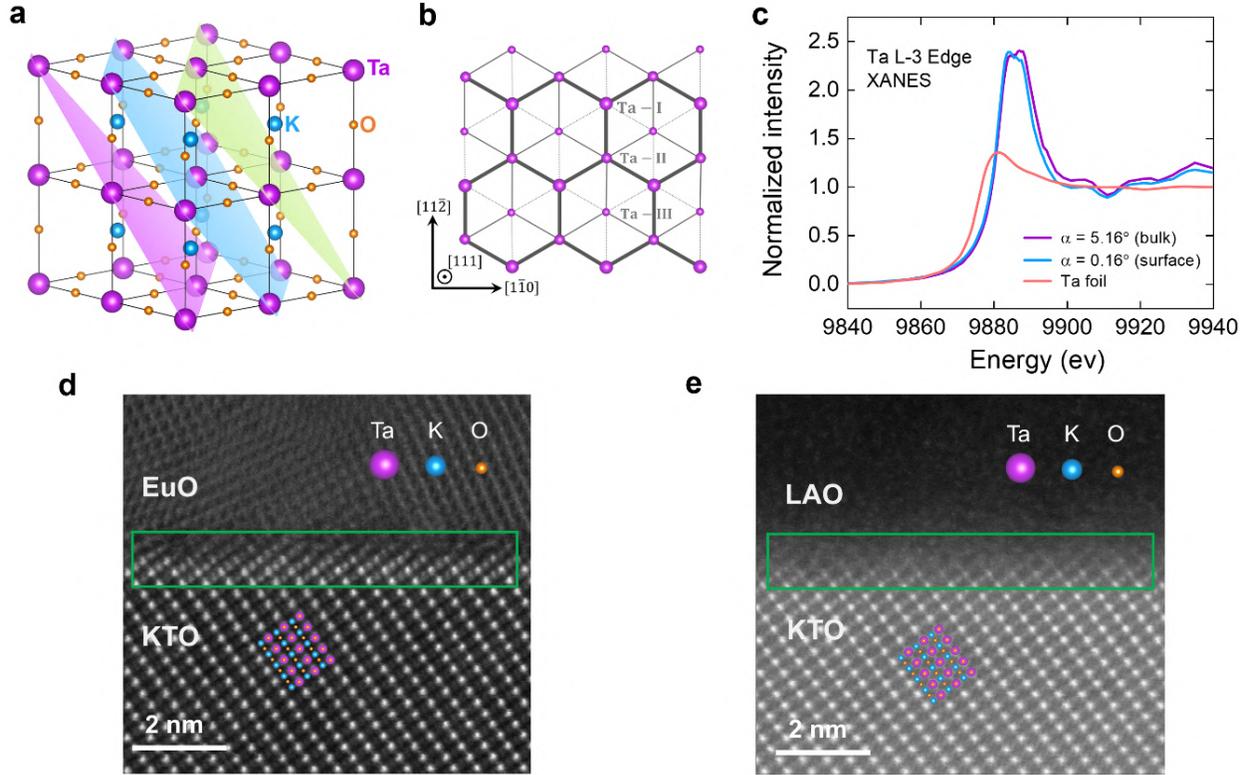

**Fig. 1. Schematics of KTO (111) surface, XANES and STEM characterizations.** (**a**) $KTaO_3$ lattice structure. The relative sizes of the ions are chosen to emphasize the Ta atoms. The three adjacent (111) planes containing $Ta^{5+}$ ions are colored in light purple, blue and green, respectively. (**b**) Distribution of $Ta^{5+}$ ions on the KTO (111) surface viewed along the [111] crystal axis. $Ta^{5+}$ ions are shown with progressively smaller sizes in the three adjacent (111) planes, which are labeled as Ta – I, Ta – II and Ta – III, respectively. Solid lines between $Ta^{5+}$ ions indicate the relative distance or coupling strength – with thicker lines representing stronger couplings, giving rise to a buckled honeycomb lattice for the first bilayer comprising Ta - I and Ta - II sites. (**c**) XANES data from sample EuO/KTO(111)_4 at the Ta *L*-edge. The KTO near the surface is only slightly reduced, with Ta valence in the surface region close to its bulk value (5+). XANES data for pure Ta is also shown for comparison. (**d**, **e**) STEM images of the EuO/KTO (111) (**d**) and LAO/KTO (111) interfaces (**e**), looking down the [1$\bar{1}$0] direction. Green box indicates the region near the interface.

Here, we report upon the observation of two-dimensional superconductivity in electron gases formed at oxide-insulator/KTO (111) interfaces. We have measured $T_c$ values as high as 2.2 K, and this value can be tuned by varying carrier density. The superconductivity is highly dependent on the crystallographic orientation of the surface on which the 2DEG is realized. That is, while the superconductivity found at the $KTaO_3$ (111) interface is robust, no superconductivity is observed in 2DEGs with similar doping levels at the $KTaO_3$ (001) interface. Recently, 2DEGs realized on the KTO (001) surface via deposition of an oxide capping layer were found to have relatively high mobilities (*28*) compared to LAO/STO, and subsequently Shubnikov de Haas oscillations were observed at low temperatures (*29*). However, no superconductivity has been reported in these or other (001) oriented samples, barring the initial discovery (*8*). Measurements of the upper critical field and current-voltage (*I-V*) characteristics confirm that the superconductivity at the oxide-insulator/KTO (111) interfaces is two-dimensional. Furthermore, a large in-plane resistive anisotropy emerges in the vicinity of the superconducting state in a series of samples with varying mobilities and carrier densities. The temperature and magnetic field dependences of this state suggest that the anisotropy is a result of the formation of a 'stripe' like phase of spatially modulated superconductivity.

**Characterization of EuO/KTO and LAO/KTO interfaces**

The 2DEGs on KTO studied in this work were prepared by growing an overlayer of EuO with molecular beam epitaxy (MBE), or LAO using pulsed laser deposition (PLD) (details in Supplementary Materials). We characterized the chemical oxidation state of the EuO/KTO (111) interface using X-ray absorption near-edge spectroscopy (XANES) at the Ta-$L_3$ edge (9.88 keV). XANES measurements were carried out adopting two different angles of incidence, one close to grazing incidence (0.16°) that probed only the first few nanometers of the KTO near the EuO/KTO interface, and another at a higher angle (5.16°) that probed the bulk of the KTO. As shown in Fig. 1c, these measurements indicate that the Ta oxidation state near the surface is very close to that in the bulk ($Ta^{5+}$), and subtle differences in the enhanced white-line feature and the absorption edge suggest that the interfacial Ta is only slightly reduced. In particular, none of our measurements show any evidence for metallic Ta, which is known to be superconducting in ultrathin films (*30*). Shown in Figs. 1d and e are scanning transmission electron microscopy (STEM) measurements of our samples, which indicate abrupt interfaces between an annealed KTO (111) surface and the polycrystalline EuO or amorphous LAO overlayer, as confirmed by X-ray diffraction measurements. Using aberration-corrected high-resolution TEM and scanning TEM, we find there are oxygen vacancies near the EuO/KTO (111) interface, and Eu substitution on the K sites within the first three unit-cells (Fig. S1), both of which can dope the interfacial KTO with electrons. Similar doping effects are also observed at the LAO/KTO (111), EuO/KTO (001) and LAO/KTO (001) interfaces (Figs. S2, S3).

**Superconductivity in 2DEG at the KTO (111) interfaces**

Figure 2a shows that the sheet resistance $R_S$ of the 2DEG in both EuO/KTO (111) and EuO/KTO (001) interfaces display metallic behavior from 300 K to 4 K. As the temperature is lowered further, the EuO/KTO (111) samples show superconductivity, which is presented in Fig. 2b for four samples. These samples were grown at different temperatures and oxygen pressures, which result in different carrier densities and mobilities. Further details of the sample growth and properties are presented in the Materials and Methods and Table S1 in the Supplementary Materials. The value of $T_c$, determined using half of the normal state resistance $R_N$, reaches as high as 2.2 K in sample EuO/KTO(111)_1. To investigate whether the EuO overlayer is necessary for superconductivity, we also prepared 2DEGs using LAO overlayers. As shown in Fig. 1c, superconductivity with $T_c > 1.1$ K is also observed in LAO/KTO (111) samples. This demonstrates that a specific overlayer of the 2DEG is not required for the emergence of superconductivity.

We have also performed measurements on 2DEGs formed at (001) oriented KTO interfaces. Figure 2d shows the measurement results for a series of (001) samples including both EuO and LAO overlayers. The charge carrier densities of these samples range from $5.8 \times 10^{13}$ cm$^{-2}$ to $1.1 \times 10^{14}$ cm$^{-2}$, covering the same range of densities as the (111) samples shown in Fig. 2b and c, though slightly lower than the $n_{2D}$ values in Ref. (*8*). Unlike the (111) samples, no superconductivity is observed in these (001) samples down to 25 mK. This crystallographic orientation dependent interfacial superconductivity is in sharp contrast with 2DEGs at STO interfaces, where the superconductivity occurs for all orientations (*10, 22, 31*), with a much lower $T_c$ of around 200 mK. STEM characterization of the (001) interfaces shows similar interfacial chemical composition as the (111) samples, which suggests that the electronic structure in the KTO (111) surface may play a role in inducing strong Cooper pairing among the interfacial charge carriers, though this point needs further investigation.

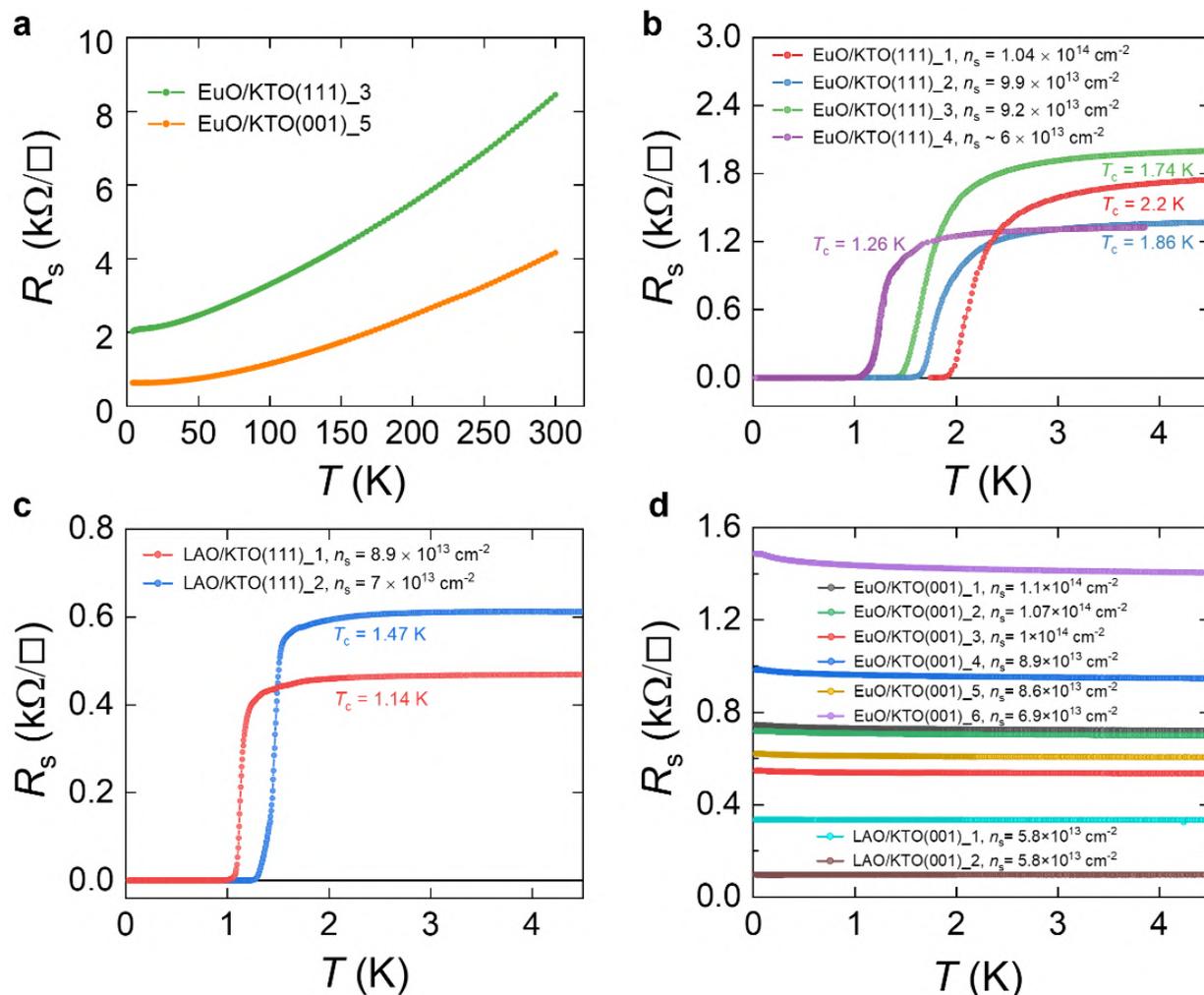

**Fig. 2. Transport measurements of 2DEGs formed at different KTO interfaces.** (**a**) Metallic temperature dependence of the sheet resistance of EuO/KTO (111) and (001) samples measured from 300 K to 4 K. (**b**) Measurement at lower temperatures shows superconducting transitions in EuO/KTO (111) samples (current along $[11\bar{2}]$) with varying carrier densities, which are determined from Hall measurement at $T$ = 10 K. The carrier density in EuO/KTO(111)_4 is estimated from growth temperature. (**c**) Similar measurements on LAO/KTO (111) samples also show superconductivity. (**d**) No superconductivity is observed in samples with (001) oriented KTO interfaces with overlayers of either EuO or LAO down to 25 mK. The range of the carrier density is similar to those of the (111) oriented samples shown in (**b**) and (**c**).

## 2D nature of the superconductivity

The superconductivity at the KTO (111) interface is two-dimensional. This is borne out by upper critical fields $B_{c\perp}$ and $B_{c\|}$ measured perpendicular and parallel to the KTO (111) interface, defined as the field at which the resistance rises to half of $R_N$. Figure 3a and b show the magnetic field dependence of the $R_S$ measured on sample EuO/KTO(111)_3. The temperature dependence of $B_{c\perp}$ is well described by the Ginzburg-Landau (G-L) theory which yields a linearized equation: $B_{c\perp} = \frac{\Phi_o}{2\pi \xi_{GL}^2}\left(1 - \frac{T}{T_c}\right)$, where $\Phi_o$ is the magnetic flux quantum and $\xi_{GL}$ is the G-L coherence length. The linear fit to $B_{c\perp}(T)$ in Fig. 3c gives $\xi_{GL} \approx 13$ nm at $T = 0$ K. The extrapolated value of $B_{c\perp}(0) \approx 1.8$ T is > $10^3$ times higher than the critical field observed in ionic-liquid gated KTO (001) devices (8). Further, we find that the critical fields are highly anisotropic. For fields applied in the sample plane, $B_{c\|}$ is much higher than $B_{c\perp}$ with $\frac{B_{c\|}}{B_{c\perp}} > 10$ at $T = 0.6$ K (Fig. 3c). The temperature dependence of $B_{c\|}$ shows a characteristic square root dependence as expected from Tinkham's model (32): $B_{c\|}(T) = \frac{\Phi_o\sqrt{12}}{2\pi d \xi_{GL}}\sqrt{1 - \frac{T}{T_c}}$, where $d$ is the effective thickness of the superconducting state. From the fit to $B_{c\|}(T)$, we obtain $d \approx 5.1$ nm, smaller than $\xi_{GL}$, consistent with 2D superconductivity.

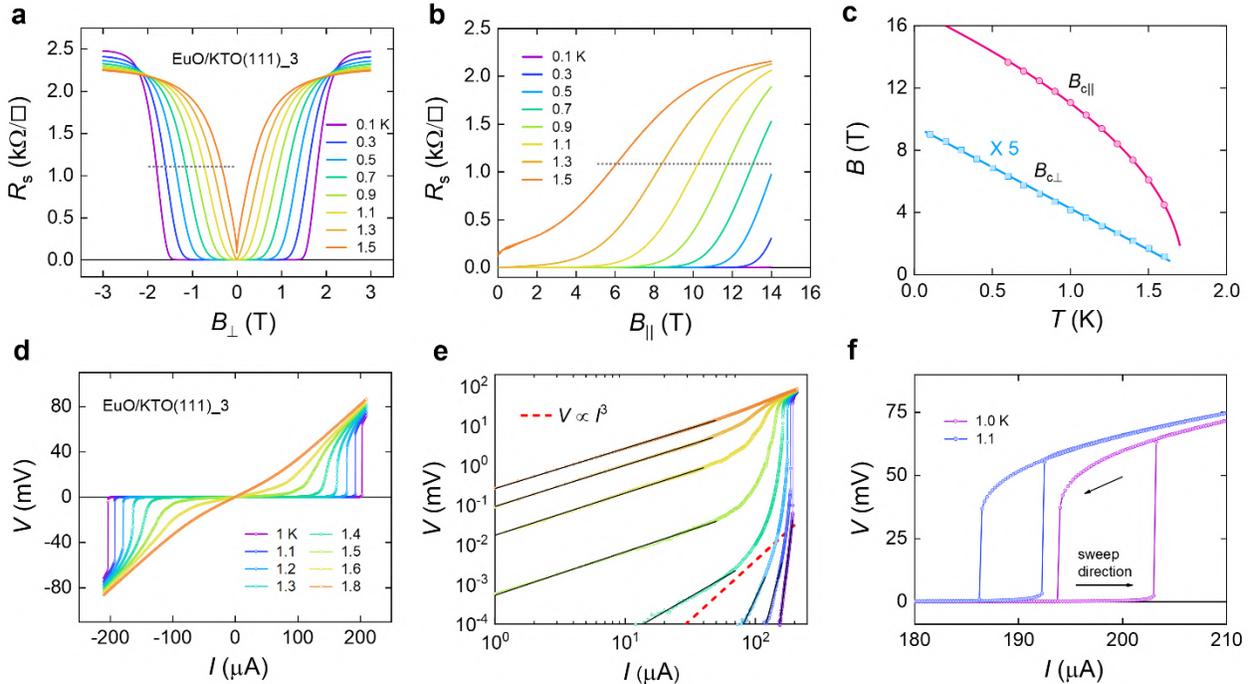

**Fig. 3. Critical field and current-voltage measurements on sample EuO/KTO(111)_3. (a, b)** Sheet resistance measured at different temperatures as a function of the out-of-plane and in-plane magnetic fields, respectively. (**c**) Temperature dependence of critical fields, which are determined at the half height of the normal-state resistance (dotted horizontal line in (**a**) and (**b**)). (**d**) *I-V* curves measured at different temperatures. (**e**) *I-V* curves plotted on a logarithmic scale using same color codes as in (**d**). Black solid lines are linear fit to the data. Red dashed line is *V*

$\propto I^3$, which is used to infer the BKT transition temperature. (**f**) Hysteresis of *I-V* curves near the critical current measured at temperatures below $T_{c0}$.

We note that $B_{c\|}(0) \approx 17.0$ T extrapolated from the fit is substantially larger than the paramagnetic pair-breaking field $B_P = \Delta_0/\sqrt{2}\mu_B \approx 3.2$ T based on BCS theory in the weak-coupling limit (*33, 34*), where $\Delta_0 \approx 1.76 k_B T_c$ is the zero-temperature superconducting gap. High values of $B_{c\|}$ in excess of $B_P$ can be realized in the presence of spin-orbit coupling due to elastic scattering (*34, 35*) in the limit where the mean free path $l_{mfp} \ll \xi_{GL}$, though other spin-relaxation mechanisms may apply even in the clean limit in the presence of broken inversion symmetry and high spin-orbit coupling (*36*). Using a single-band model, our estimates suggest $l_{mfp} \sim 0.4$ to 0.7 times $\xi_{GL}$, which is intermediate between the clean and dirty limits. Enhanced $B_{c\|}$ values have also been reported in superconductors that are believed to have a spin-triplet *p*-wave order parameter, though these were thought to be unstable in the presence of disorder. However, recent theoretical developments indicate that in the presence of strong spin-orbit coupling, criteria that might apply to KTO (111) interfaces such as broken inversion symmetry (*37*), spin-momentum locking (*38*) or multi-orbital superconducting states (*39*) can stabilize unconventional superconductivity against disorder scattering. Thus, at this time, the mechanism for realizing $B_{c\|}$ values in excess of $B_P$ at KTO (111) interfaces remains an open question. In Fig. 3a, we also observe an approximate crossing point near $B_\perp = 2.1$ T, which implies the existence of a nearly temperature-independent separatrix in $R_s$ vs. *T* traces that show either a positive or negative temperature coefficient of resistance upon cooling, depending on magnetic field (see also Fig. S4). Such data may be interpreted in terms of a quantum phase transition between a 2D superconductor and a correlated metal (*40*).

The superconducting state in the EuO/KTO sample shows a robust critical current. The measured *I-V* characteristics at different temperatures are shown in Fig. 3d, where the critical current $I_c$ approaches values close to 200 µA at *T* = 1 K. We estimate a lower bound for two-dimensional critical current density, $K_c > 500$ µA cm$^{-1}$ at 1K, which is more than 700 times larger than the $K_c = 0.7$ µA cm$^{-1}$ reported for devices on ionic-liquid gated KTO (001) surfaces (*8*). As the temperature is raised to values near $T_c$, we observe the gradual onset of a resistive state at low currents. In Fig. 3e we plot the *I-V* on a log-log scale, and observe that the slope of the *I-V* characteristics evolves smoothly from the normal Ohmic state $V \propto I$ towards a steeper power law $V \propto I^\alpha$ as superconductivity sets in at lower temperatures. This evolution may be interpreted in terms of a Berezinskii–Kosterlitz–Thouless (BKT) transition in a 2D superconductor (*41*), where the onset of a non-linear *I-V* characteristic in the superconducting state is due to the current driven unbinding of vortex anti-vortex pairs that are created by thermal fluctuations at finite temperatures. This causes the exponent $\alpha$ to smoothly evolve from a value of 1 above and near the onset of superconductivity to higher values as the temperature is lowered (*42, 43*), with the $\alpha = 3$ at $T_{BKT}$. Red dashed line in Fig. 3f corresponds to a $I^3$ dependence, which indicates that the $T_{BKT}$ of this sample EuO/KTO(111)_3 is between 1.2 K and 1.3 K, which is slightly below the zero-resistance $T_{c0}$ of about 1.34 K. At temperature below $T_{c0}$, we observe hysteresis in the *I-V* curves near $I_c$ (Fig. 3f), which suggests that the 2D superconductivity is inhomogeneous, with superconducting regions joined by weak links (*44*).

**Emergence of an anisotropic 'stripe' phase near the superconducting state**

In higher mobility EuO/KTO (111) samples, we observe the emergence of a new phase near the superconducting state, as revealed by a giant in-plane anisotropy in sheet resistance. Shown in Fig. 4a is the measurement in sample EuO/KTO(111)_5, which has a relatively high mobility of about 279 cm$^2$ V$^{-1}$ s$^{-1}$ and a low charge carrier density of 3.5 × 10$^{13}$ cm$^{-2}$ at $T$ = 10 K. The measurements are carried out in a van der Pauw geometry as shown in Fig. 4b. As temperature is lowered below about 2.2 K (Fig. 4a), the resistance *increases* by almost 50% for current along the [1$\bar{1}$0] crystal axis, while it *decreases* by over 50% for current flowing in the [11$\bar{2}$] direction. The in-plane anisotropy observed here is far stronger than that observed at LAO/STO (111) 2DEGs (*45, 46*). The formation of this anisotropy over the entire sample suggests the emergence of a new phase (*47*) which breaks rotational symmetry over macroscopic length scales (Fig. S5). The observed anisotropy persists over a broad temperature range from 2.2 K down to about 0.7 K, which is shown with a green background in Fig. 4a. At still lower temperatures, the resistance in both the [1$\bar{1}$0] and [11$\bar{2}$] current directions drop rapidly to zero, and the superconducting state is obtained (light blue region).

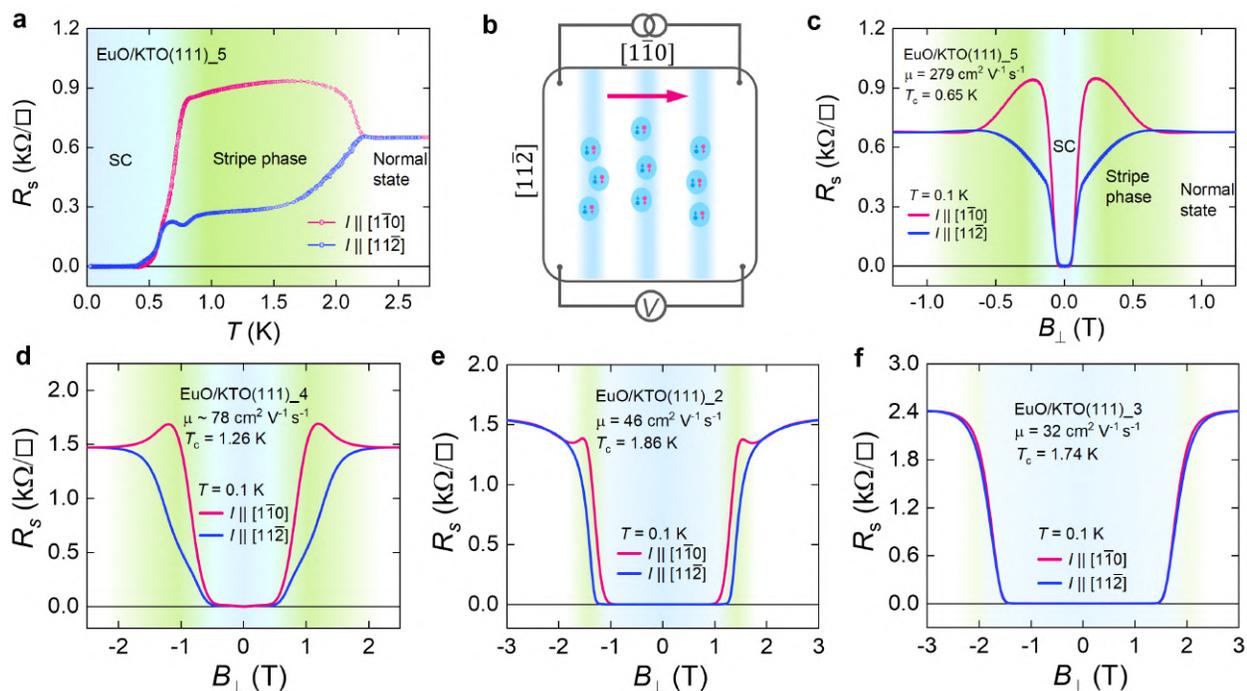

**Fig. 4. Stripe phase in samples with varying mobilities.** (**a**) Sheet resistance of sample EuO/KTO(111)_5 measured with electric current along [1$\bar{1}$0] (red) and [11$\bar{2}$] (blue) crystal axes under zero field. The light blue and green region indicate superconducting (SC) and 'stripe' state, respectively. (**b**) Illustration of the measurement geometry for the case of current along [1$\bar{1}$0] direction perpendicular to the stripes. These stripes may be composed of Cooper pairs, which are shown as light blue bubbles. (**c**)-(**f**) Magnetic field dependence of the sheet resistance measured along both current directions at $T$ = 0.1 K in samples with decreasing mobilities. The stripe phase

is revealed around the critical field (green region). Note that EuO/KTO(111)_2 has a higher $T_c$ than EuO/KTO(111)_3, but also shows a more prominent stripe phase.

The increase in resistance along the $[1\bar{1}0]$ direction upon lowering temperature from the normal state is similar to observations in 2D granular superconductors (*48*). This occurs due to the gapping out of quasiparticles by a superconducting gap (*49*), which inhibits transport between weakly coupled superconducting regions. Global superconductivity is restored at still lower temperatures presumably via Josephson coupling between these regions. The decrease in resistance along $[11\bar{2}]$ at the same temperature implies stronger coupling between the superconducting regions in that direction. Such an anisotropy would naturally occur if the superconducting regions organized themselves into 'stripes' as shown in Fig. 4b. Recently, pre-formed superconducting puddles have also been suggested in the LAO/STO system, where macroscopic 2D superconductivity is obtained when phase coherence develops between them (*50*).

The magnetic field dependence of sheet resistance provides further evidence for an anisotropic 'stripe' like superconductivity. Figure 4c shows the resistance measured in the same EuO/KTO(111)_5 sample in an out-of-plane magnetic field at $T = 0.1$ K, which is well below $T_{c0}$. As the magnetic field is increased to the critical field, there is a sharp rise in the resistance along both directions of current. The initial high/low resistance values along $[1\bar{1}0]/[11\bar{2}]$ in the resistive state are approximately equal to those in the 'stripe' phase shown in Fig. 4a, obtained by just raising the temperature to above $T_{c0}$. This suggests that the 'stripe' phase is obtained by a suppression of global superconductivity upon the application of a magnetic field. By increasing the magnetic field further, the resistance *decreases* along $[1\bar{1}0]$ and increases along $[11\bar{2}]$, as the 'stripe' phase is driven into the normal state at $B_\perp \gtrsim 0.8$ T, where the resistance anisotropy also disappears, presumably due to a more complete suppression of the superconducting gap within the stripes. The decrease in resistance with magnetic field for transport along $[1\bar{1}0]$ due to the suppression of the superconducting gap is also a signature of 'granular' superconductivity in that direction.

Figures 4d-f show magnetotransport measurements in other higher-carrier-density EuO/KTO (111) samples, which have lower mobilities than that of EuO/KTO(001)_5. The resistance anisotropy, with the same orientation relative to the crystal axes, is observed in all of these samples to varying degrees around the critical field. The magnitude of the transport anisotropy measured relative to the normal-state value becomes progressively smaller as the mobility of the sample decreases. In sample EuO/KTO(111)_3, which has the lowest mobility among these samples, the transport is nearly isotropic as shown in Fig. 4f. This observation suggest that the 'stripe' phase depends on the level of disorder in the sample. Furthermore, the 'stripe' phase seems to not compete with the superconductivity. For example, sample EuO/KTO(111)_2 clearly shows 'stripe' like anisotropy (Fig. 4e), but it also has higher $T_c$ than that of EuO/KTO(111)_3 (see Fig. 2b), which does not show the anisotropy (Fig. 4f). We also note that the transport in LAO/KTO (111) is mostly isotropic when an out-of-plane field is applied (Fig. S6), however a large anisotropy is observed upon applying an *in-plane* magnetic field approaching $B_{c\parallel}$. This suggests that the 'stripe' phase may be influenced by coupling between the magnetic EuO overlayer and the superconductivity in EuO/KTO (111) samples, or by an in-plane magnetic field as in LAO/KTO (111).

**Discussion**

The 2D superconductivity observed in the KTO (111) interfaces shows substantially higher $T_c$ than that observed in the widely studied LAO/STO system, and increases monotonically with $n_{2D}$. These higher $T_c$ values making fundamental studies and device applications of superconducting 2DEGs using crystalline interfaces more readily accessible. The origin of the superconductivity is still an open question. Possible links between superconductivity and the large spin-orbit interaction in KTO (111) 2DEGs, as has been suggested in LAO/STO (111) (*51, 52*), could be explored in future work.

When global superconductivity in the KTO (111) surface is suppressed by either temperature or a magnetic field, transport measurements reveal a 'stripe' phase which produces large anisotropic transport in the (111) plane. Notably, the directions for the high/low resistance states in the vicinity of superconductivity are oriented along the same crystal axes in the KTO and STO (111) interfaces (*46*). We observe no dependence on thermal cycling across the onset temperature of the 'stripe' phase (Fig. S7), as has been observed in LAO/STO (*53*), where these effects were due to the formation of structural domain walls. In addition, the 'stripe' phase shows strong dependences on magnetic field and scattering of charge carriers by disorder, which suggest an electronic origin. A possible explanation for the observed anisotropy may be a spatially modulated or inhomogeneous superconductivity. This can happen for example in a Fulde-Ferrel-Larkin-Ovchinnikov (FFLO) state, where a spin-split Fermi surface due to a magnetic proximity effect or an external magnetic field results in Cooper pairs with non-zero total crystal momentum (*37, 54*), whose density is then periodically modulated in real space. It is also possible that the superconductivity at KTO (111) interfaces is inherently composed of a spatially modulated density of Cooper pairs or a pair density-wave (PDW) (*55*). Further experiments, including probes of the spatial structure of superconductivity, will be helpful for elucidating the nature of superconductivity and resistance anisotropy that we observe.

**Acknowledgments:** We thank Peter Littlewood, Ivar Martin and Michael Norman for discussions. We thank Chengjun Sun for the reference XANES spectral of Ta metal foil, and M. Eblen-Zayas for discussions about EuO growth. **Funding:** All work at Argonne was supported by the US Department of Energy, Office of Science, Basic Energy Sciences, Materials Sciences and Engineering Division. The use of facilities at the Center for Nanoscale Materials and the Advanced Photon Source, both Office of Science user facilities, was supported by the US Department of Energy, Basic Energy Sciences under Contract No. DE-AC02-06CH11357. The support provided by China Scholarship Council (CSC) during a visit of Xi Yan to Argonne National Laboratory is acknowledged. J. S. acknowledges the support of the National Natural Science Foundation of China (No. 11934016). J.-M. Z is supported by the Energy & Biosciences Institute through the EBI-Shell program. The work at Peking University was supported by National Natural Science Foundation of China (Grant No. 11974025); **Author contributions:** Superconductivity was discovered in EuO/KTO (111) and LAO/KTO (111) samples by C. L. with assistance from X. Y., T. B. S, J. S. J, X. Z. and D. J.. EuO/KTO samples presented here were grown using MBE based techniques at Peking University by Y. M. and W. H. (see Fig. S8 and Table S1) and at Argonne by C. L. with assistance from J. P. and A. B.; LAO/KTO samples were made at Argonne using PLD by X. Y. with assistance from D. D. F.. All low-temperature transport measurements were planned by C. L. and A. B.. Dilution fridge based magnetotransport measurements were carried out in a 14-Tesla magnet by C. L., T. B. S. and J. S. J. at the Materials Science Division (MSD), Argonne, and in a split-coil (5-1-1 T) magnet by C. L., X. Z. and D. J. at the Center for Nanoscale Materials (CNM), Argonne. Transport measurements down to 1.75 K were carried out by C. L. in a Quantum Design Physical Property Measurement System cryostat at the CNM, Argonne with assistance from B. F.. Samples from Peking University were measured in collaboration with X. Y. and J. S.. X-ray diffraction, both bulk and surface, and XANES measurement at the Ta *L*-edge were carried out by H. Z. and X. Y. at the Advanced Photon Source at Argonne, with assistance from D. D. F.. TEM sample preparation and aberration-corrected HRTEM measurements were carried out by Y. L. and J. W. at CNM, Argonne. STEM and STEM-EELS measurements were carried out by H.-W. H. and J.-M. Z. at UIUC. The paper was written by C. L. and A. B. with inputs from all co-authors. A. B. supervised the project; **Competing interests:** Authors declare no competing interests; and **Data and materials availability:** The data that support the findings of this study are available from the corresponding author upon reasonable request.

**Supplementary Materials:**

Materials and Methods, Figures S1-S8, Table S1

# Supplementary Materials for

# Discovery of two-dimensional anisotropic superconductivity at KTaO3 (111) interfaces

Changjiang Liu[1*†], Xi Yan[1,2,5 †], Dafei Jin[3], Yang Ma[6], Haw-Wen Hsiao[4], Yulin Lin[3], Terence M. Bretz-Sullivan[1], Xianjing Zhou[3], John Pearson[1], Brandon Fisher[3], J. Samuel Jiang[1], Wei Han[6], Jian-Min Zuo[4], Jianguo Wen[3], Dillon D. Fong[1], Jirong Sun[5], Hua Zhou[2], Anand Bhattacharya[1*].

[1]Materials Science Division, Argonne National Laboratory, Lemont, IL 60439, USA.

[2]Advanced Photon Source, Argonne National Laboratory, Lemont, IL 60439, USA.

[3]Nanoscale Science and Technology Division, Argonne National Laboratory, Lemont, IL 60439, USA.

[4]Department of Materials Science and Engineering, University of Illinois at Urbana-Champaign, Urbana, Illinois 61801, USA.

[5]Beijing National Laboratory for Condensed Matter & Institute of Physics, Chinese Academy of Sciences, Beijing 100190, People's Republic of China.

[6]International Centre for Quantum Materials, School of Physics, Peking University, Beijing 100871, People's Republic of China.

*Correspondence to anand@anl.gov or changjiang.liu@anl.gov

†These authors contributed equally.

**This PDF file includes:**

Materials and Methods

Figs. S1 to S8

Table S1

**Materials and Methods**

Growth of EuO/KTO samples

The EuO/KTO samples were grown using molecular beam epitaxy. The growth chamber has a base pressure of about $2 \times 10^{-10}$ torr. Prior to the growth, the KTO substrate was annealed in vacuum at 600 °C for about 30 mins. The substrate temperature was then reduced to the growth temperatures, which are listed in Extended Data Fig. 3c. The Eu was heated in an effusion cell to a deposition temperature of about 455 °C, and is evaporated at stable rate of about 0.08 Å/s as measured by a quartz crystal microbalance (QCM) at the substrate position. Ultra-high purity oxygen was delivered into the chamber through a valve which allowed for fine control of oxygen partial pressure in the range from $2 \times 10^{-10}$ torr to $2 \times 10^{-8}$ torr. The Eu atoms have a low sticking coefficient to the substrate at high temperatures ($\geq$ 450 °C), unless oxygen is delivered into the chamber. The first few layers of EuO was grown by depositing pure Eu under vacuum which getters oxygen and promotes the formation of oxygen vacancies near the KTO surface. Oxygen was then delivered into the chamber with a partial pressure that gradually increased from $5 \times 10^{-10}$ torr to about $1.5 \times 10^{-8}$ over the course of growth. After growth, a 3-nm germanium or 5-nm MgO was grown to protect sample from further oxidation in air. For samples grown at low temperatures, such as EuO/KTO(111)_5, oxygen was not required because the sticking coefficient of Eu was sufficient. After being exposed to air, the overlayer presumably undergoes complete oxidation which was sufficient to protect the 2DEG on the KTO surface.

Growth of LAO/KTO samples

Amorphous LAO thin films were prepared on the as received 5 × 5 mm KTO substrates using pulsed laser deposition (PLD) method. The substrates were pre-annealed in situ at 500 °C in an oxygen partial pressure of $10^{-6}$ torr for 30 mins before growth. During the deposition, temperature and pressure were kept the same and a 248-nm KrF excimer laser was used with 2 Hz repetition rate and 1.5 J/cm$^2$ laser fluence. After growth, the samples were cooled down to room temperature at a rate of 5 °C/min under the same oxygen pressure as during growth. A commercially available 10 × 10 mm single crystal LAO (001) wafer was used as the target. The surface morphologies of the substrates and films were atomically flat, investigated by atomic force microscopy (AFM).

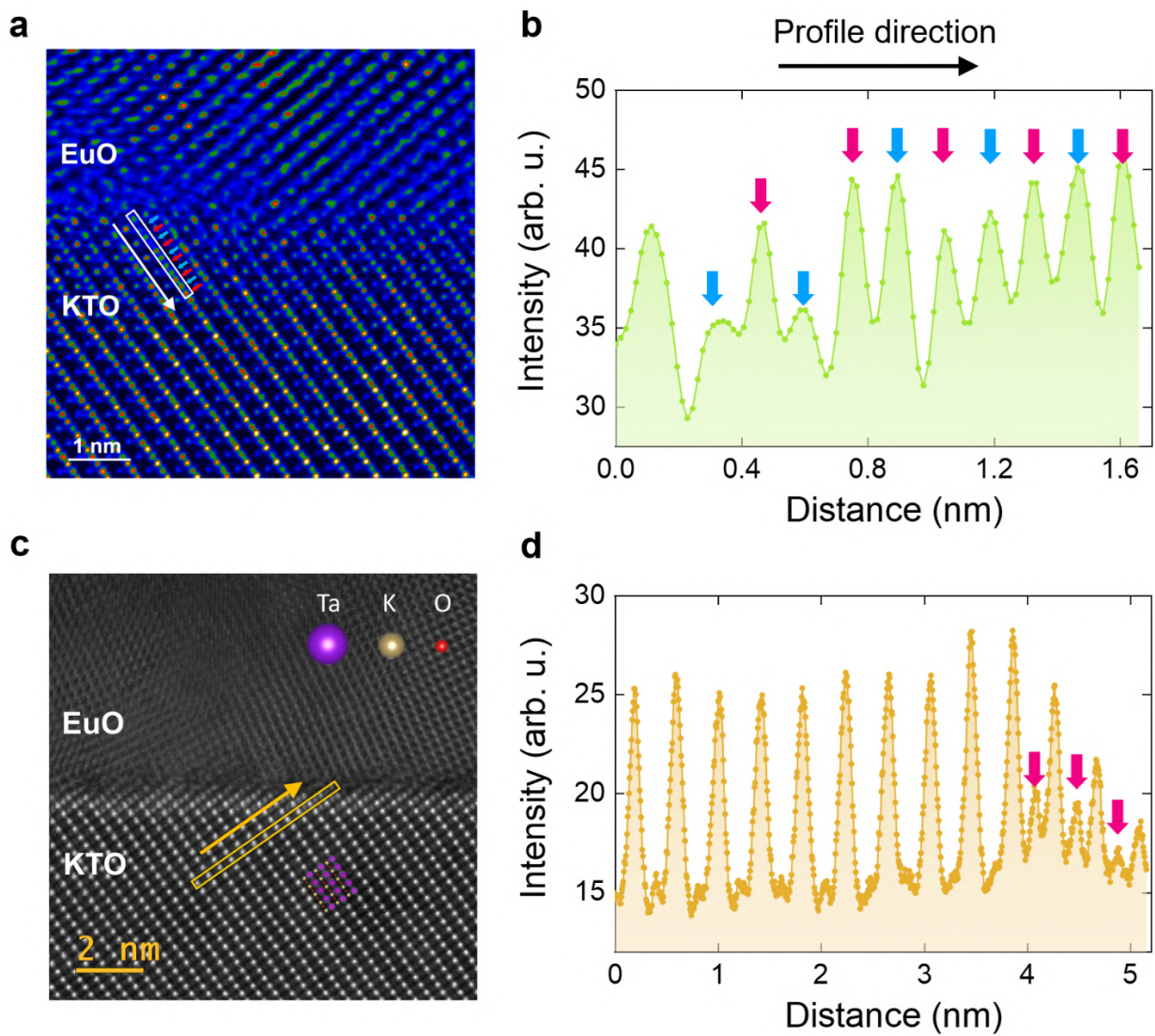

**Fig. S1. Characterization of interfacial compositions on sample EuO/KTO(111)_4**. (**a**) Aberration-corrected high-resolution scanning transmission electron microscopy (STEM) of the EuO/KTO (111) interface in annular bright field (ABF). The white box in (**a**) is a TaO$_2$ plane, where Ta and O columns are marked by red and blue arrows, respectively. (**b**) Intensity profile of the box in (**a**). The intensity of two oxygen columns near the interface is lower than in the bulk, indicating oxygen vacancies in KTO near the interface. (**c**) A high angle annular dark field (HAADF) image of the EuO/KTO sample acquired by using an aberration-corrected scanning transmission electron microscope (Themis Z, Thermo-Fisher Scientific), operated at 300kV. (**d**) The line profile in the region indicated by the orange box in (**c**), in the direction towards the interface. The potassium sites near the interface (indicated by red arrows) show stronger intensity than in the bulk of the substrate, suggesting that some potassium atoms are replaced by europium atoms.

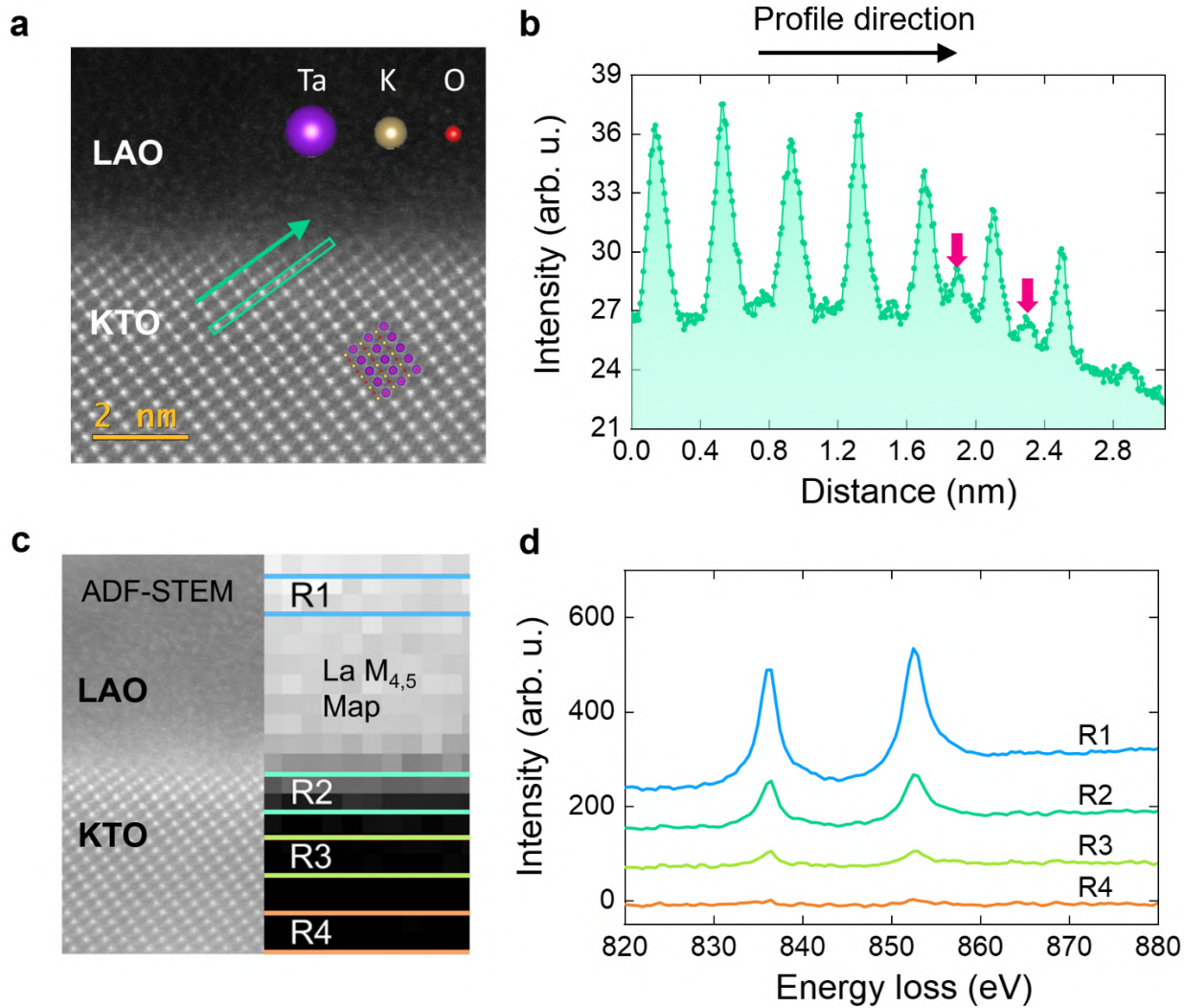

**Fig. S2. Characterization of interfacial compositions on a LAO/KTO (111) sample.** (**a**) A HAADF image of a LAO/KTO (111) sample acquired by using an aberration-corrected scanning transmission electron microscope (Themis Z, Thermo-Fisher Scientific), operated at 300 kV. (**b**) The line profile inside the region indicated by the green box in (**a**), in the direction towards the interface. The potassium sites near the interface (indicated by red arrows) show higher intensity than in the bulk, suggesting that some potassium atoms may be replaced by lanthanum atoms. (**c**) Lanthanum intensity map from an EELS scan across the interface (20 by 20 points with a step size of 0.5 nm). (**d**) Area average spectra of lanthanum $M_{4,5}$ edges in regions R1, R2, R3 and R4. The clear EELS peaks in R2 suggests La doping in KTO near the interface with LAO. The amount of La drops significantly in the region of R4.

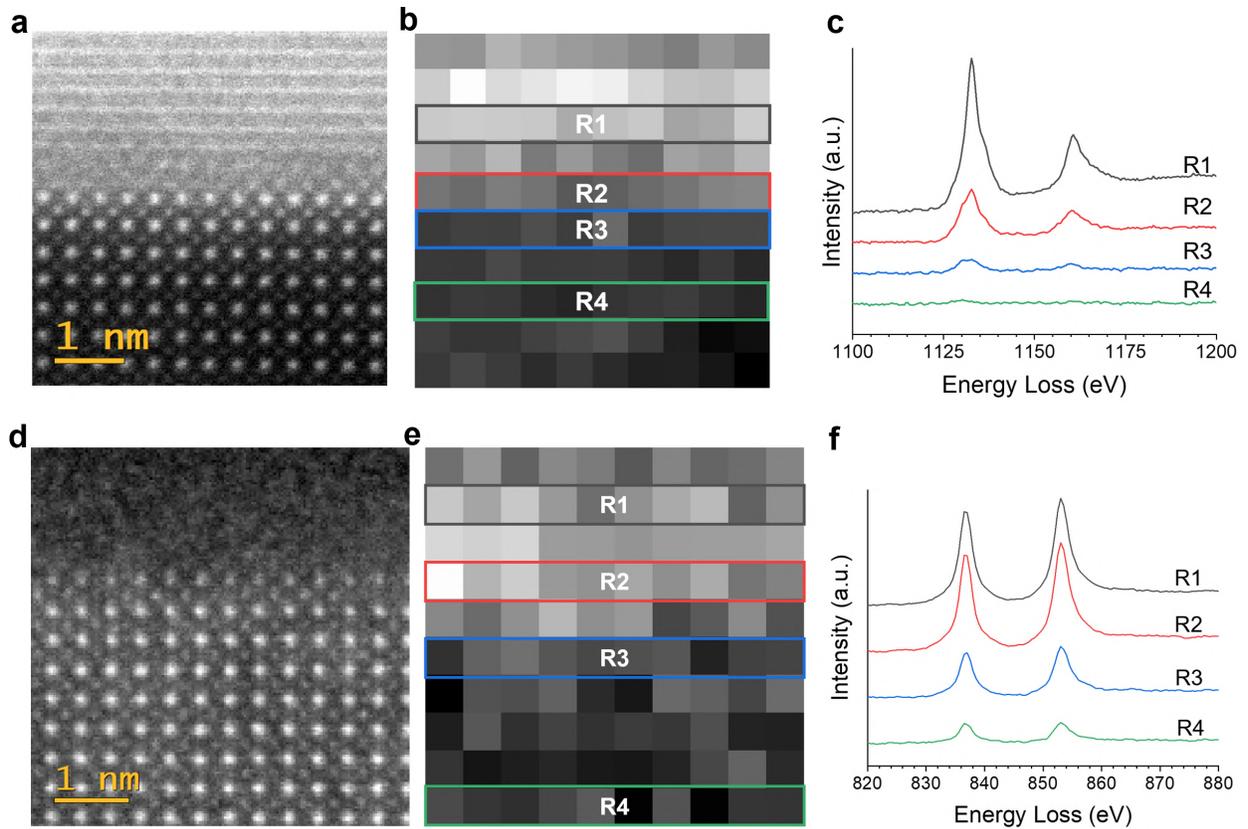

**Fig. S3. Characterization of interfacial composition on (001) oriented samples.** (**a**) A HAADF image of EuO/KTO(001)_5. (**b**), (**c**) Area average spectra of Eu $M_{4,5}$ edges in regions R1 (EuO) and R2, R3 and R4. The clear Eu EELS peaks in R2-R3 suggest Eu doping in KTO near the interface with EuO, extending into approximately 3-unit cell of KTO. (**d**) A HAADF image of a LAO/KTO (001). (**e**), (**f**) Area average spectra of lanthanum $M_{4,5}$ edges in regions R1 (LAO), R2, R3 and R4. The clear EELS peaks in R2-R4 suggest La doping in KTO near the interface with LAO. The amount of La drops by R4, though not as much as in the (111) interface. The La diffuses beyond 8-unit cells of KTO in the (001) direction.

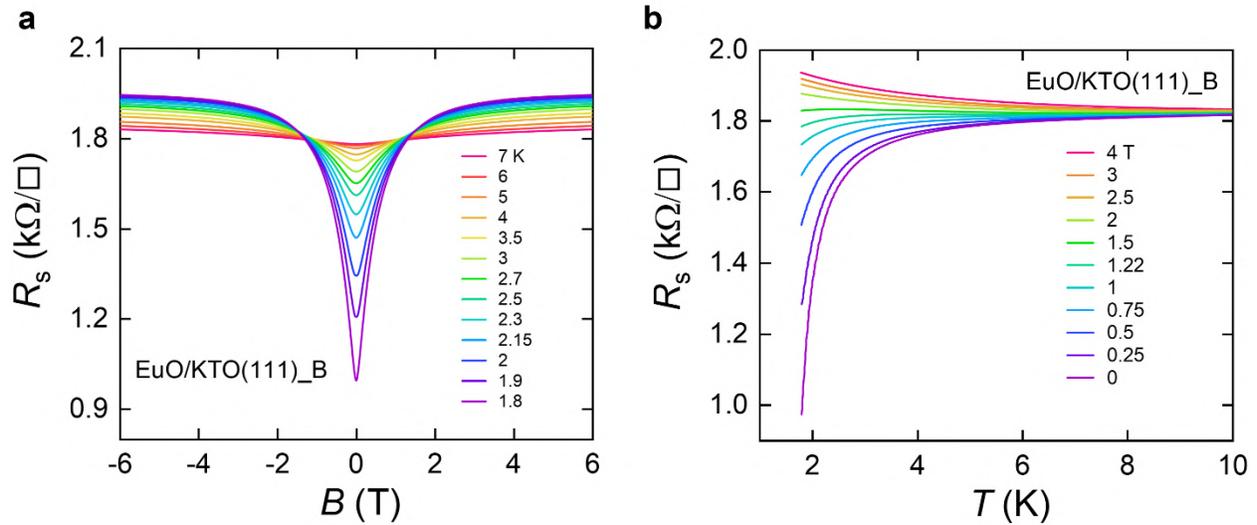

**Fig. S4. Transport measurements on EuO/KTO(111)_3 at higher temperatures.** (**a**) Magnetic field dependence of the sheet resistance measured from 1.8 K to 7 K. At low field, the low resistance state is due to superconducting fluctuations which are gradually suppressed with increasing magnetic field, resulting in a positive magnetoresistance. Signs of superconducting fluctuations persist to about 7 K, much higher than the superconducting transition $T_c$. (**b**) Temperature dependence of the sheet resistance measured under different magnetic fields. As magnetic field increases, the superconducting behavior - large positive temperature coefficient of resistance (TCR), gradually evolves into a semi-insulating behavior (negative TCR). This may be understood as a magnetic-field induced transition between a 2D superconductor and a correlated metal. At field ~ 1.5 T, the resistance remains constant over a broad temperature range. We note that the sheet resistance of the 2DEG drifts up slowly over time. Data shown here were obtained shortly after the sample was grown. Data shown in Fig. 3 of the main text were taken after one month. The increase in resistance during that time period was about 15%.

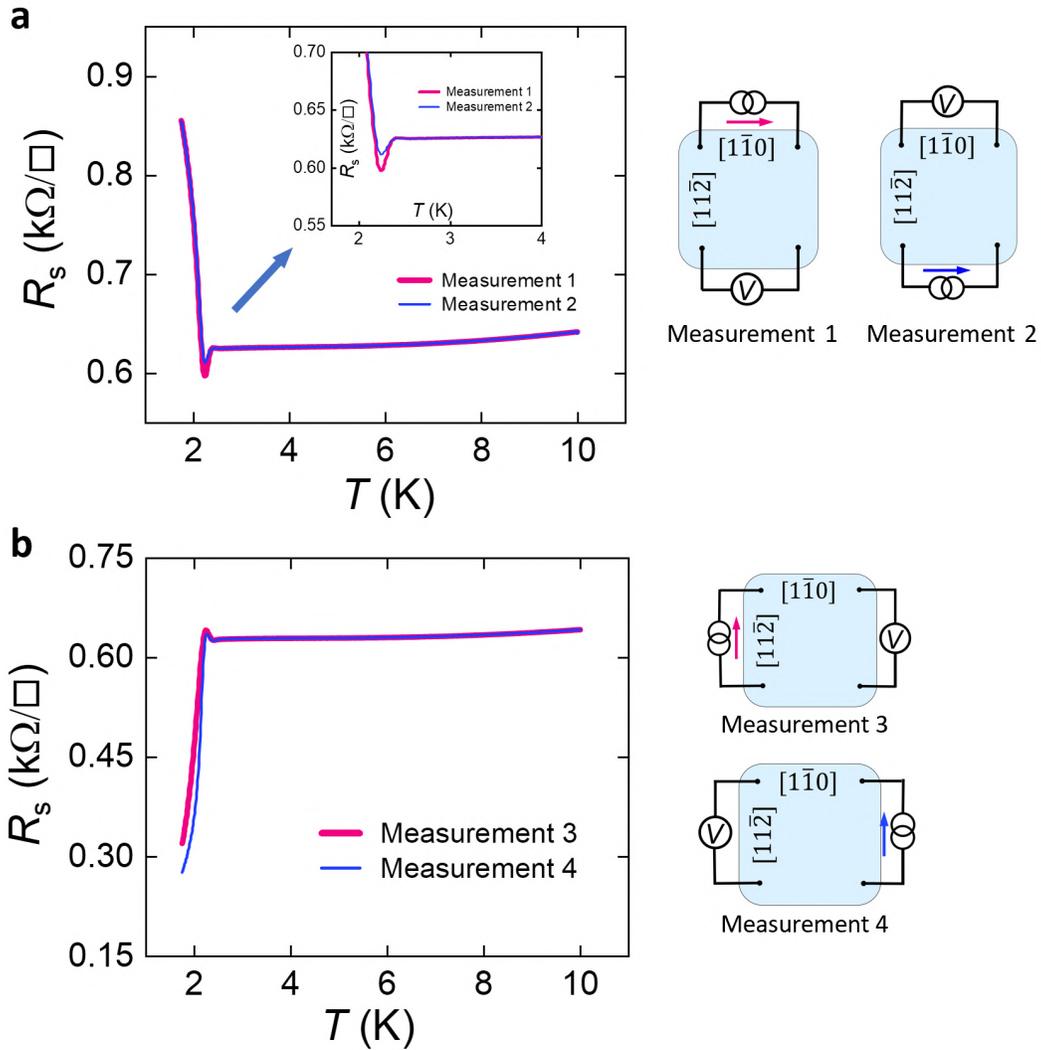

**Fig. S5. Homogeneity of the sample EuO/KTO(001)_5.** Current and voltage leads were switched in the van der Pauw measurement (shown on the right) and a similar anisotropic resistance was measured on both sides of the sample for current along $[1\bar{1}0]$ (**a**) and along $[11\bar{2}]$ (**b**). The small difference in the sheet resistance (blue and red data) in the stripe phase may be due to the high sensitivity of the stripes to local disorder in the sample. Notably, the blue and red data show the same resistance value at $T > 2.2$ K, which indicates that the resistivity of the sample is homogeneous in the normal state.

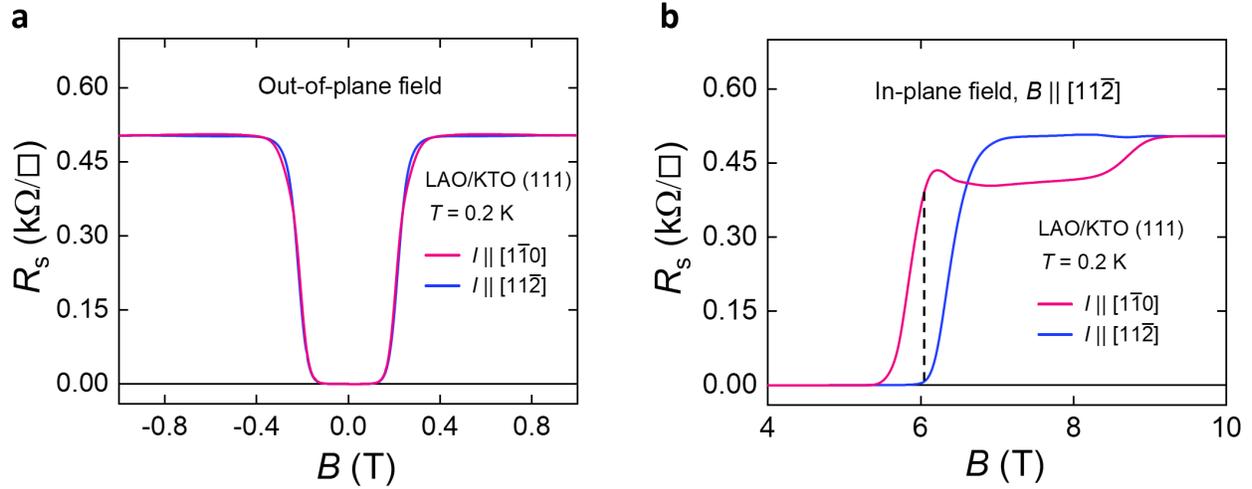

**Fig. S6. Stripe phase in the LAO/KTO (111) sample.** (**a**) The resistance measured as a function of magnetic field is mostly the same between current along $[1\bar{1}0]$ (red) and along $[11\bar{2}]$ (blue), when the field is applied along the out-of-plane direction. (**b**) Large anisotropy in resistance is observed near the in-plane critical field $B_{c\parallel} \sim 6\ T$ as indicated by the vertical dashed line. Note that the high/low resistance state near the critical field corresponds to the same crystal axis as observed in the EuO/KTO (111) samples.

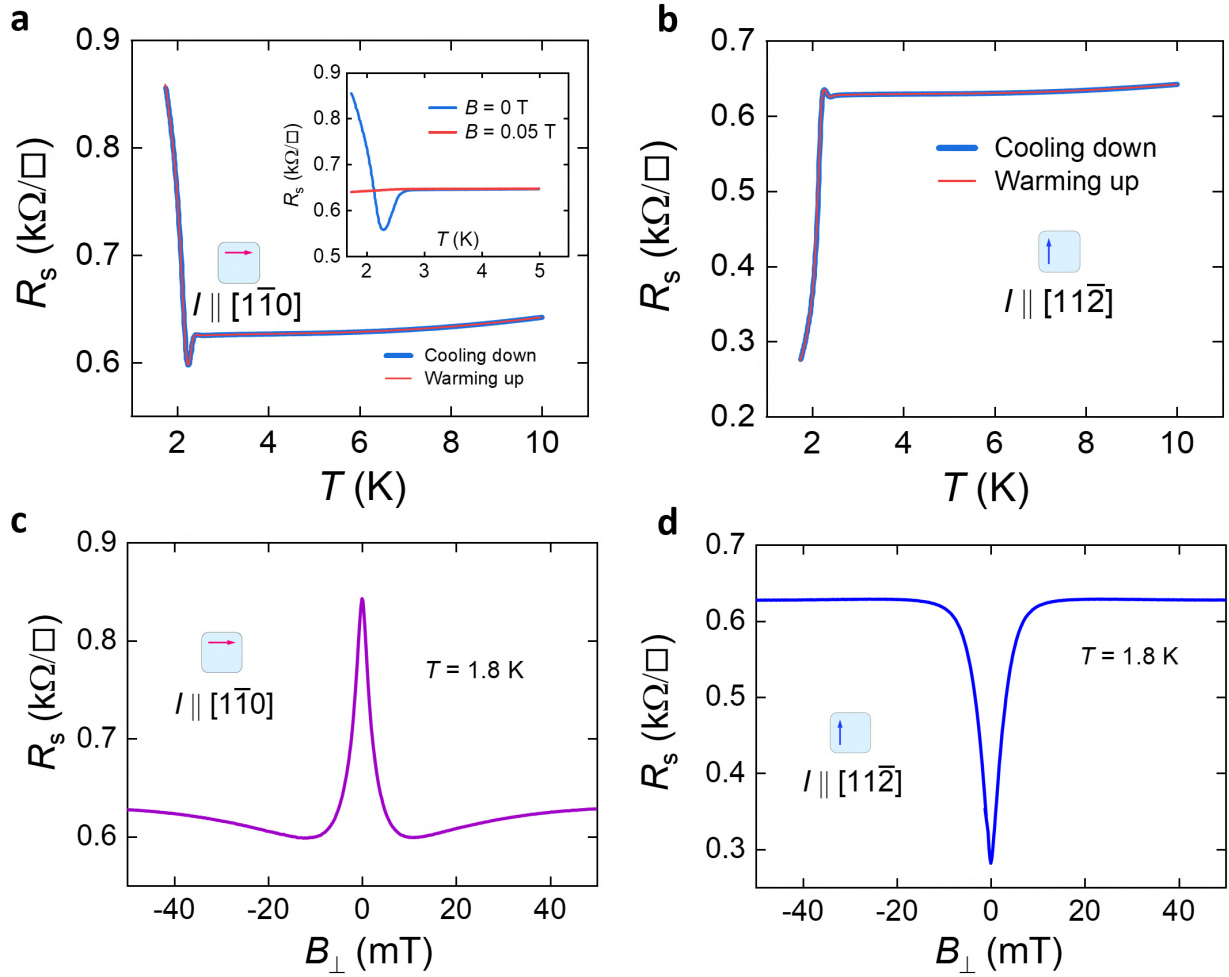

**Fig. S7. Effects of thermal cycling and magnetic field on resistance anisotropy.** Resistance anisotropy in sample EuO/KTO(111)_5 is insensitive to thermal cycling, with measurements of performed along $[1\bar{1}0]$ (**a**) and $[11\bar{2}]$ (**b**). Inset of (**a**) shows that the resistance rise along the $[1\bar{1}0]$ direction is suppressed to below 1.75 K upon an application of an out-of-plane magnetic field of 0.05 T. (**c**), (**d**) Magnetic field dependence of the sheet resistance at $T = 1.8$ K for current along $[1\bar{1}0]$ and $[11\bar{2}]$ crystal axis, respectively. Note the sharp negative and positive magnetoresistance in the two current directions.

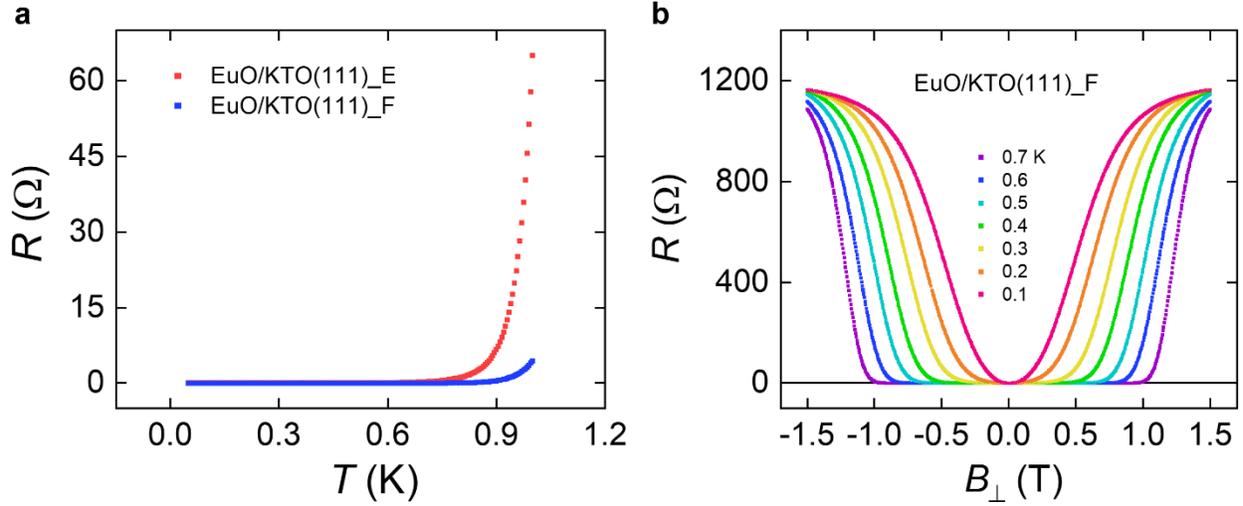

**Fig. S8. Superconducting transition and magnetotransport measurements on samples from Peking University.** (**a**) Superconductivity observed in sample EuO/KTO(111)_E and F, which were grown at Peking University. The measurements were performed at temperatures up to 1 K. The zero-resistance state transition temperatures are $T_{c0} \approx 0.8$ K and 0.6 K for EuO/KTO(111)_E and F, respectively. (**b**) Out-of-plane critical field measurement on sample EuO/KTO(111)_F. The Ginzburg-Landau coherence length obtained from this measurement is $\xi_{GL} \approx 15.4$ nm. Besides these two EuO/KTO (111) samples.

| Samples | Growth $T$ | $n_s$ ($10^{13}$ cm$^{-2}$) | $R_s$ (kΩ/□), 10 K | $\mu$ (cm$^2$ V$^{-1}$ s$^{-1}$) | $T_c$ |
|---|---|---|---|---|---|
| EuO/KTO(001)_1 | 500 °C | 11 | 0.72 | 79 | NA |
| EuO/KTO(001)_2 | 500 °C | 10.7 | 0.7 | 83 | NA |
| EuO/KTO(001)_3 | 500 °C | 10 | 0.54 | 116 | NA |
| EuO/KTO(001)_4 | 500 °C | 8.9 | 0.95 | 74 | NA |
| EuO/KTO(001)_5 | 500 °C | 8.6 | 0.64 | 114 | NA |
| EuO/KTO(001)_6 | 500 °C | 6.9 | 1.4 | 65 | NA |
| | | | | | |
| EuO/KTO(111)_1 | 500 °C | 10.4 | 1.74 | 34 | 2.2 K |
| EuO/KTO(111)_2 | 550 °C | 9.9 | 1.36 | 46 | 1.86 K |
| EuO/KTO(111)_3 | 500 °C | 9.2 | 2 | 34 | 1.74 K |
| EuO/KTO(111)_4 | 450 °C | ~ 6 | 1.38 | ~ 75 | 1.26 K |
| EuO/KTO(111)_5 | 375 °C | 3.5 | 0.64 | 279 | 0.65 K |
| | | | | | |
| LAO/KTO(001)_1 | 500 °C | 5.8 | 0.096 | 1122 | NA |
| LAO/KTO(001)_2 | 500 °C | 5.8 | 0.333 | 323 | NA |
| | | | | | |
| LAO/KTO(111)_1 | 500 °C | 8.9 | 0.468 | 150 | 1.14 K |
| LAO/KTO(111)_2 | 500 °C | 7 | 0.611 | 146 | 1.47 K |

**Table S1. Summary of sample properties.** Sample EuO/KTO(001)_1, _2, _3, _4, and _6 were grown at Peking University. The other samples were grown at Argonne.